# Convection in rotating flows with simultaneous imposition of radial and vertical temperature gradients

Ayan Kumar Banerjee[1], Siddhesh Tirodkar[2], Amitabh Bhattacharya[1,2], Sridhar Balasubramanian[1,2]

[1]Department of Mechanical Engineering, IIT Bombay, India

[2]Interdisciplinary program in climate studies, IIT Bombay, India

Email: ayanbanerjee1@gmail.com

**Abstract**

Laboratory experiments were conducted to study heat transport characteristics in a non-homogeneously heated fluid annulus subjected to rotation along the vertical axis ($z$). The non-homogeneous heating was obtained by imposing radial and vertical temperature gradient ($\Delta T$). The parameter range for this study was Rayleigh number, $Ra=2.43\times10^8$-$3.66\times10^8$, and Taylor number, $Ta=6.45\times10^8$-$27\times10^8$. The working fluid was water with a Prandtl number, $Pr=7$. Heat transport was measured for varying rotation rates ($\Omega$) for fixed values of $\Delta T$. The Nusselt number, $Nu$, plotted as a function of $Ta$ distinctly showed the effect of rotation on heat transport. In general, $Nu$ was found to have a larger value for non-rotating convection. This could mean an interplay of columnar plumes and baroclinic wave in our system as also evident from temperature measurements. Laser based imaging at a single vertical plane also showed evidence of such flow structure.

## 1 Introduction

Thermal convection in the presence of rotation is an important fluid dynamic problem commonly encountered in geophysical and astrophysical flows. In general, two variants of this problem have been studied extensively, (i) Convective dynamics in non-rotating and rotating fluid with vertical temperature gradient (classical Rayleigh-Benard convection (RBC)), studied by Rossby (1969), Liu and Ecke (2009), King (2012) among many others and (ii) Convective dynamics in rotating flow with radial temperature gradient (classical Hide-Mason experiment) studied by Hide (1953), Wordsworth (2008) and Vincze (2013) among many others. However, the convective dynamics of a rotating fluid due to simultaneous imposition of radial and vertical temperature gradient is poorly understood. Hignett (1981), Park (1998), Sheard (2011), Hussam (2014) and few others tried to fill this gap by studying a system called rotating horizontal convection (RHC) where radially varying constant temperature is provided on the bottom plate. Although our present study has some similarity to RHC, our system has vertical temperature gradient only along the periphery. Such system is helpful in understanding meridional heat transport in ocean. In general, study of RHC focuses on understanding the boundary layer dynamics. But the present study also focuses on bulk heat transfer characteristics, relating it to flow structure.

## 2. Experimental set up and different parameters

A rotating table was built in-house for conducting experiments to study convective dynamics in a differentially heated rotating system. The table was controlled using Variable Frequency Drive (VFD) and was equipped with measurement devices. The set up comprises of a cylindrical tank with a flat bottom made of acrylic with an aluminum periphery of 5 mm thickness. A smaller diameter copper cylinder is placed at the centre on the bottom surface concentrically whose geometric centre coincides with the axis of the rotation. A chiller was used to maintain a steady cold temperature on the inner cylinder. The aluminum periphery on the bottom plate is heated from below using an electrical heater whose temperature was regulated by a PID-DAQ system. Such an arrangement ensures a thermal gradient in both radial and vertical directions. In the present set up, radial temperature difference decreases along the direction of rotation (+z-axis) in non-rotating condition as shown in Figure1. Distilled water was used as working fluid to fill the annular gap between the two cylinders with depth, $d$=12 cm. Two more DAQ systems were also employed for acquisition of temperature time series by thermocouples. The distributions of thermocouples were: 10 thermocouples at one fixed azimuthal position but at different vertical heights ($z$) on the outer cylinder and 8 thermocouples on the inner cylinder. A laser light source was mounted on the rotating frame to create a vertical sheet to image on different vertical planes in the flow field. PIV particles of 300μm diameter were used for flow visualization. A camera was attached to rotating table side ways to record flow pattern in the vertical plane.

The main governing non-dimensional numbers are: Rayleigh number ($Ra$) which is the ratio of buoyancy forces to the product of momentum and thermal diffusivity, and proportional to radial temperature difference, and the Taylor number ($Ta$) which is ratio of the Coriolis force to viscous

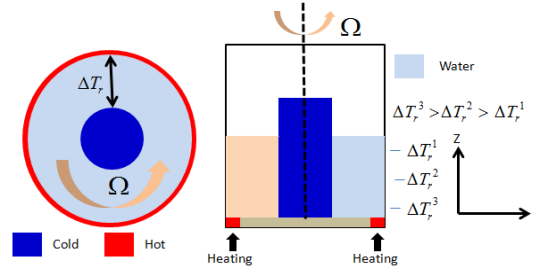

Figure.1: Schematic of experimental set up; Orange rectangular plane on left side of the tank denotes vertical plane which is simulated in ANSYS-Fluent

forces, and characterizes impact of rotation. Moreover in the differentially heated rotating annulus experiment, aspect ratio ($\Gamma$) becomes very important, hence both *Ta* and *Ra* are modified to include aspect ratio ($\Gamma$). Thus the complete set of governing non-dimensional numbers are:

$$\Gamma = \frac{d}{b-a} \quad \widetilde{Ta} = Ta \times \frac{1}{\Gamma} \quad \widetilde{Ra} = Ra \times \frac{1}{\Gamma} \quad Pr = \frac{\nu}{\kappa} \qquad (1)$$

which can be written as:

$$\widetilde{Ta} = \frac{4\Omega^2 (b-a)^5}{\nu^2 d} \qquad \widetilde{Ra} = \frac{g\alpha \Delta T\ (b-a)^4}{d\nu\kappa} \qquad (2)$$

where, $g$ is the acceleration of gravity (m/s$^2$), α is the volumetric thermal expansion coefficient (K$^{-1}$), $\Delta T$ is the temperature difference (°C), ν is the kinematic viscosity (m$^2$/s), κ is the thermal diffusivity (m$^2$/s), and $\Omega$ is the angular rotation rate (rotation per second), $\Gamma$ is aspect ratio, $b$ is the outer radius (m), $a$ is the inner radius (m), $d$ is the depth (m) of the working fluid in the annulus. Henceforth, we will use modified Taylor number and Rayleigh number (i.e. $\widetilde{Ta}$, $\widetilde{Ra}$).

## 3. Results and discussions

Here we present the experimental results on heat transport in the presence and absence of rotation collected using multiple thermocouples embedded at different vertical positions on both inner and outer wall. In addition to experiments, ANSYS Fluent was used to simulate flow dynamics in a 2-D axisymmetric plane. The in-built governing equations in ANSYS were used to model the flow.

### 3.1. Experimental results

Chandrasekhar (1953) in his pioneering work showed the effect of rotation on convective dynamics. Since then many others such as Hide (1953), Rossby (1969), Park (1998), Liu and Ecke (2009), King (2012), Wordsworth (2008), Vincze (2013), Sheard (2011), Hussam (2014) etc. studied different kinds of rotating convective dynamics. They studied effects of differential temperature imposed (buoyancy) and rotation on heat flux transported in terms of non-dimensional numbers. Amount of heat transport and thermal convection in a system is characterized by a non-dimensional number, Nusselt number ($Nu$). $Nu$ is defined as the ratio between the heat flux in the presence of convection to the conductive heat flux that would exist without fluid motion:

$$Nu = \frac{Q_{convection}}{Q_{diffusion}} = \frac{(Q-Q_{loss})(b-a)}{KA\,\Delta T} \qquad (3)$$

where, $Q_{loss}$ is the sum of background losses due to conduction from the side wall through plexi glass and from insulator (glass wool). $K$ (W/mK) is the thermal conductivity of fluid, $A$ is the surface area over which heat flux was provided, $(b-a)$ is the radial gap over which differential temperature was imposed. These losses were calculated and were included in the $Nu$ measurements for better accuracy. Influence of temperature difference and rotation on heat transports was found by investigating effect of $Ra$ and $Ta$ on $Nu$. Different scaling behavior between $Nu$, $Ra$ and $Ta$ are known at different regimes for rotating RBC and rotating horizontal convection. The present work investigates the variation of $Nu$ with varying $Ra$, $Ta$ for the system taken here, hence tries to establish scaling behavior. The list of parameters studied are shown in Table1. Experiments were conducted for varying values of *Ta* for two different *ΔT* e.g. for one particular radial temperature gap (*ΔT*), a range of *Ta* was covered. All the experiments were done for aspect ratio 1.0.

Table.1: Parameter range for this study

| # Run | $\Delta T$ | $Ra$ | $Ta$ |
|---|---|---|---|
| 1 | 10° C | 2.437×10$^8$ | 0 |
| 2 | 10° C | 2.437×10$^8$ | 6.45 ×10$^8$ |
| 3 | 10° C | 2.437×10$^8$ | 9.44 ×10$^8$ |
| 4 | 10° C | 2.437×10$^8$ | 1.5 ×10$^9$ |
| 5 | 10° C | 2.437×10$^8$ | 2.7 ×10$^9$ |
| 6 | 15° C | 3.66 ×10$^8$ | 0 |
| 7 | 15° C | 3.66 ×10$^8$ | 6.45 ×10$^8$ |
| 8 | 15° C | 3.66 ×10$^8$ | 9.44 ×10$^8$ |
| 9 | 15° C | 3.66 ×10$^8$ | 1.5 ×10$^9$ |
| 10 | 15° C | 3.66 ×10$^8$ | 2.7 ×10$^9$ |

The heat transfer in thermal convective flows is characterized by the Nusselt number, *Nu*. For the set of experimental runs shown in Table 1, *Nu* was calculated for varying *Ta* characterizing different rotation rates for two different *ΔT*. It was found that *Nu* significantly decreases for rotating condition compared to non-rotating convection, possibly indicating rotation induced constraint. In other words, convection in non-rotating case is three dimensional contrary to rotating case where heat is transported through vertical columnar plume near the outer wall. Hence radial temperature gap is comparatively more, resulting in lower *Nu* compared to non-rotating case.

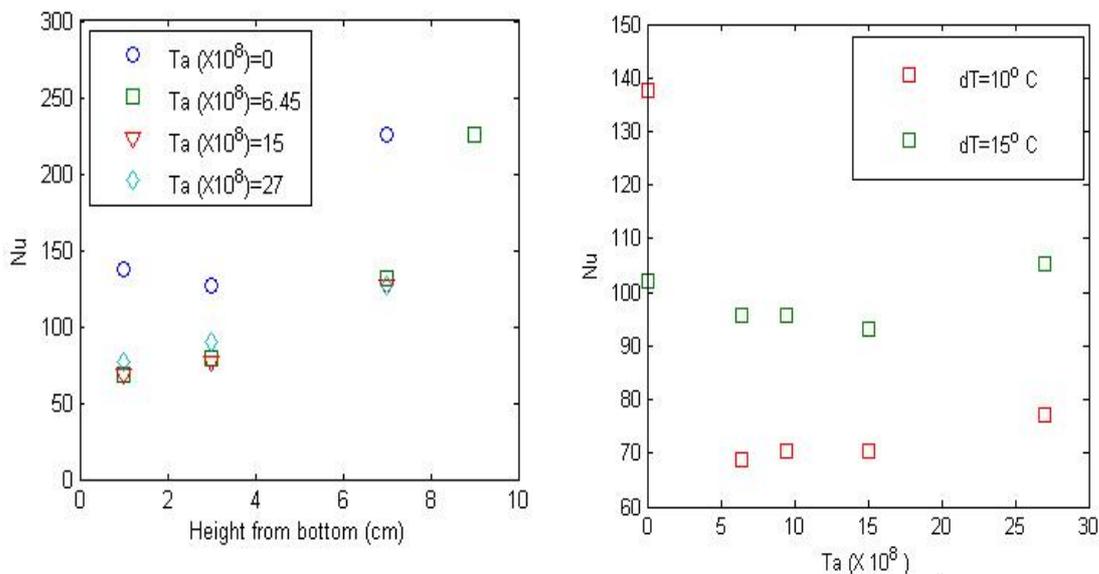

Figure.2: Left- Variation of *Nu* with elevation. Plot is done for *Ra*=3.66 ×$10^8$; Right- Variation of *Nu* with varying Taylor numbers. Data are plotted for *ΔT* =15 °C and *ΔT* =10 °C

However, for rotating condition *Nu* increases with the increase of rotation due to more efficient heat transport by Taylor columns at higher rotation. Furthermore, it is counter-intuitive that *Nu* is lower for higher *ΔT* in rotating condition. At the same time for no rotation (i.e. *Ta*=0) *Nu* is higher for higher *ΔT*. This issue will be addresses later on. Finally a scaling behavior between *Nu* and *Ta* will be found. Variation of *Nu* with elevation is also plotted for *Ra*=3.66 ×10$^8$. In this case *Nu* was calculated at different heights based on the *ΔT* at those heights. *Nu* was found to be increasing with elevation which is obvious due to slant constant temperature line imposed by baroclinic instability. Moreover, peripheral bottom heating does not reach to the top completely, making the upper fluid layer thermally near homogeneous. Hence *Nu* tends to be very high at high elevation but that does not necessarily mean occurrence of higher convection at higher elevation.

Temperature time series data were captured by thermocouples embedded in system for the range of parameters in Table 1. Data for *Ta*=6.45×10$^8$ and *Ra*=3.61 × 10$^8$, on outer wall at one azimuthal location with varying *z* is plotted below:

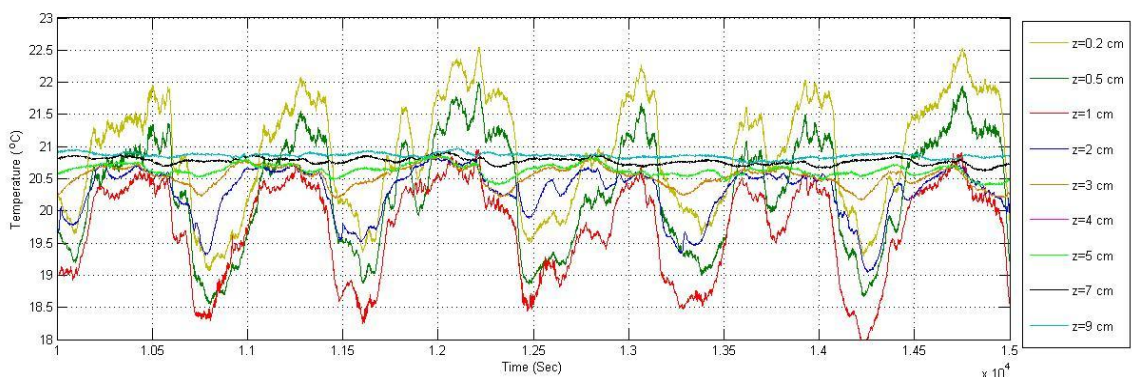

Figure 3: Temperature time series at a fixed azimuthal location on outer wall with increasing heights from bottom

Temperature data hints at the coexistence of thermal plume and baroclinic waves in the flow field. Fluctuation in the temperature decreases with elevation indicating decreasing impact of heating with elevation. Temperature profiles at a fixed azimuthal position are found to be in phase indicating existence of columnar structure aligned parallel to the axis of rotation. These plumes aid in vertical heat transport. A phase difference was found in temperature time series data taken at two azimuthally differing locations on outer wall, indicating a phase velocity associated with the thermal plumes.

Velocity field on a vertical plane at 1.5 cm distance away from inner cylinder was captured. Baroclinic waves cuts that plane while moving azimuthally. Vorticity plot shows existence of vortices in that plane, although no convective plume like structures were found as evident from temperature data. The vertical velocity (v) was found to be of the order of less than 1 mm.s$^{-1}$. Further analysis of PIV images is underway for better understanding of the flow dynamics.

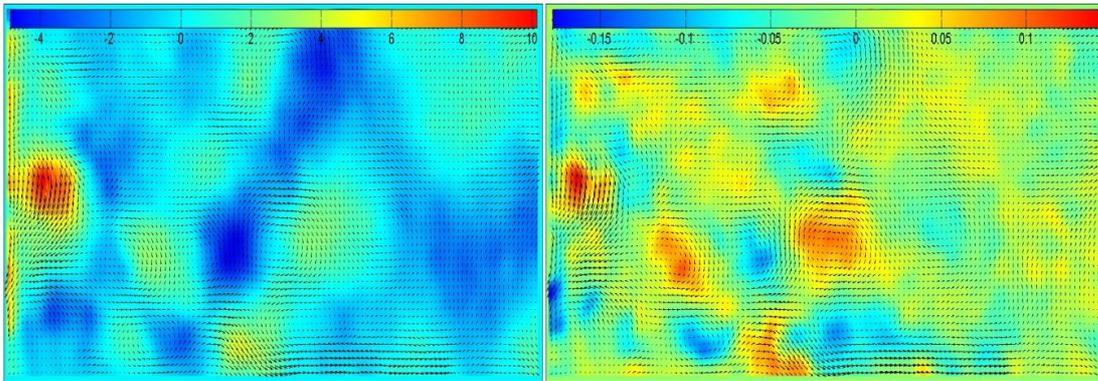

Figure 4: Left: PIV image showing v-component of velocity on a vertical plane. Velocity scale given in image is to be multiplied with 10$^{-4}$. Right: Vorticity field on the same plane.

**3.2. Simulation results**

The problem is symmetric with respect to the axis of rotation. ANSYS-Fluent was used to simulate flow dynamics on the vertical plane in between central cold cylinder and periphery for two different cases, $Ta$=0, 6.45 ×10$^8$ (Figure1). Simulations were done for constant heating from bottom (peripherally) at 303 K and constant cooling in the centre at 288 K. All other sides were kept in adiabatic condition. In stationary case, although thermal plume exists but baroclinic instability does not exist. The inclined isotherms in rotating case suggest existence of baroclinic instability. Furthermore, simulation results indicate existence of helicity in the azimuthal velocity.

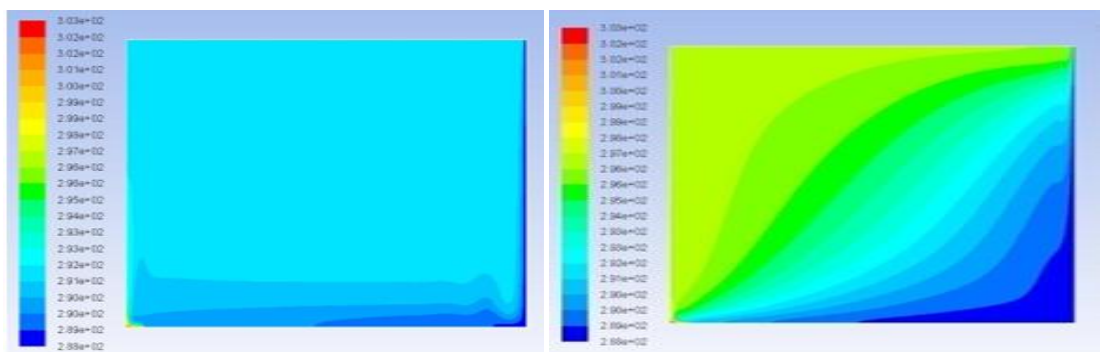

Figure 5: Comparison of simulated static temperatures for $Ta$=0 (left) and $Ta$=6.45 ×10$^8$ (right)

**Conclusion**

An experimental study of turbulent convection in rotating and non-rotating frame with tandem radial and vertical temperature gradient is presented. For the case of non-rotating convection, the Nusselt number, *Nu*, was found to exceed its value in rotating convection. Again, *Nu* is comparatively lower for higher *Ra* in rotating condition. Temperature time series measurement hints at the coexistence of thermal plume and baroclinic waves in case of rotating convection, which will be addressed further in future studies.

**References**

S. Chandrasekhar, (1953). The instability of a layer of fluid heated below and subject to Coriolis forces, in *Proc. R. Soc. Lond. A*, vol. 217, pp.306-327.

G. Ahlers, S. Grossmann, and D. Lohse, (2009) Heat transfer and large scale dynamics in turbulent Rayleigh-Benard convection.*Rev. Mod. Phys.*,vol. 81, pp. 503-537.

H. T. Rossby,(1969).A study of Benard convection with and without rotation,*J. Fluid Mech.*, vol. 36, pp. 309-335.

Y. Liu and R. E. Ecke,(2009).Heat transport measurements in turbulent rotating Rayleigh-Bénard convection,*Phys. Rev. E*, vol. 80, pp. 1-12.

E. M. King and J. M. Aurnou, (2012). Thermal evidence for Taylor columns in turbulent rotating Rayleigh-Benard convection,*Phys. Rev. E.*, vol. 85, pp. 1-11.

H. T. Rossby,(1965). On thermal convection driven by non uniform heating from below: an experimental study.*Deep-sea Research.*, vol. 12, pp. 9-16.

Y.G.Park and J.A.Whitehead,(1998). Rotating Convection Driven by Differential Bottom Heating," *J. phy. Oceanography*,vol. 19.

Larcher, Egbers,(2005). Experiments on trasition of baroclinic waves in differentially heated rotating annulus, Nonlin. Processes Geophys,vol.12, pp.1033–1041.

Vincze, M; Harlander, U; von Larcher, Thand Egbers, (2013). An experimental study of regime transitions in a differentially heated baroclinic annulus with flat and sloping bottom topographies. Nonlin. Processes Geophys.

Hignett et. al,On rotating thermal convection driven by non uniform heating from below. *J. Fluid Mech.*, vol. 109, pp. 161-187, 1981.